\documentclass[11pt]{article}
\usepackage[textwidth=15.2cm,textheight=22cm]{geometry}
\usepackage{amsmath,amssymb}
\usepackage{latexsym}
\usepackage{slashed}
\usepackage{multicol}
\usepackage{graphicx}
\usepackage{bm}
\usepackage{tikz}
\usepackage{cite}
\usepackage{tikz}
\usetikzlibrary{trees}
\usepackage{varwidth}
\usepackage{pifont}
\usepackage{cancel}
\usepackage{centernot}
\usepackage{mathtools}
\usepackage{float}

\newcommand{\xdownarrow}[1]{%
  {\left\downarrow\vbox to #1{}\right.\kern-\nulldelimiterspace}
}
\usetikzlibrary{shapes.geometric, arrows, shadows}

\tikzstyle{forces} = [rectangle, rounded corners, minimum width=2cm, minimum height=0.8cm,text centered, draw=black]
\tikzstyle{spin} = [rectangle, rounded corners, minimum width=1.3cm, minimum height=0.8cm,text centered, draw=black]
\tikzstyle{theory} = [rectangle, rounded corners, minimum width=2cm, minimum height=0.8cm,text centered, draw=gray]
\tikzstyle{arrow} = [thick,->,>=stealth]
\tikzstyle{arrow1} = [thick,<->,>=stealth]
\tikzstyle{line} = [draw, -latex']

\allowdisplaybreaks[1]

\newcommand{\be}{\begin{equation}}
\newcommand{\ee}{\end{equation}}
\newcommand{\ba}{\begin{eqnarray}}
\newcommand{\ea}{\end{eqnarray}}
\newcommand\fr[1]{\frac{1}{#1}}

\newcommand{\rom}[1]{\uppercase\expandafter{\romannumeral #1\relax}}

\def\s{\sigma}

\def\d{\delta}
\def\D{\Delta}

\def\ba{\bar A}

\def\beq{\begin{equation}}
\def\eeq{\end{equation}}

\newcommand{\nn}{\nonumber}

\newcommand{\ndt}{\noindent}

\def\bea{\begin{eqnarray}}
\def\eea{\end{eqnarray}}
\def\beas{\begin{eqnarray*}}
\def\eeas{\end{eqnarray*}}
\def\sla{\raise.15ex\hbox{$/$}\kern-.57em}

\def\d{\delta}
\def\m#1{\mathcal#1}

\def\spa#1.#2{\left\langle#1\,#2\right\rangle}
\def\spb#1.#2{\left[#1\,#2\right]}
\def\cT{{\cal T}}
\def\m{\mu}
\def\n{\nu}
\def\pa{{\partial}}
\def\cO{{\cal O}}
\def\q{\quad}
\def\qq{\quad\quad}

\def\Ra{{\Rightarrow}}
\def\c{\gamma}

\begin{document}

\begin{titlepage}
\begin{flushright}    
{\small $\,$}
\end{flushright}
\vskip 1cm
\centerline{\LARGE{\bf{Supersymmetric Yang-Mills theory in $D=6$}}}
\vskip 0.5cm
\centerline{\LARGE{\bf{without anti-commuting variables}}}
\vskip 1.5cm
\centerline{Sudarshan Ananth$^{\,\dagger}$, Hannes Malcha$^{\,\star}$, Chetan Pandey$^{\,\dagger}$ and Saurabh Pant$^{\,\dagger}$}
\vskip 0.5cm
\centerline{${\,}^\dagger$\it {Indian Institute of Science Education and Research}}
\centerline{\it {Pune 411008, India}}
\vskip 0.5cm
\centerline{${\,}^\star$\it {Max-Planck-Institut f\"ur Gravitationsphysik (Albert-Einstein-Institut)}}
\centerline {\it {Am M\"{u}hlenberg 1, 14476 Potsdam, Germany}}
\vskip 1.5cm
\centerline{\bf {Abstract}}
\vskip .5cm
Supersymmetric Yang--Mills theory is formulated in six dimensions, without the use of anti-commuting variables. This is achieved using a new Nicolai map, to third order in the coupling constant. This is the second such map in six dimensions and highlights a potential ambiguity in the formalism.
\vfill
\end{titlepage}

\section{Introduction and Notation}

\ndt Supersymmetric theories may be formulated without the use of anti-commuting variables~\cite{Nic1, Nic2}. In this approach, supersymmetric gauge theories are characterized by a Nicolai map -- a transformation of the bosonic fields such that the Jacobian determinant of the transformation exactly cancels against the product of the Matthews-Salam-Seiler (MSS)~\cite{MSS} and Faddeev-Popov determinants~\cite{FP}. The formalism avoids any use of anti-commuting objects thus offering an alternate perspective on the physics of gauge theories.
\vskip 0.3cm
\ndt The map, for Yang-Mills theory, was explicitly constructed to second order in the coupling constant in~\cite{Nic2}, refined in~\cite{DL1} and derived from a rigorous R-prescription in~\cite{L1}. It was subsequently shown~\cite{Nic3} that this construction holds in {\it {all}} the critical dimensions $D=3,4,6,10$ where supersymmetric Yang–Mills theories exist~\cite{SYM1}. The map and the framework itself were extended to third order in the coupling constant in~\cite{Nic4}. 
\vskip 0.3cm
\ndt In this paper, we present a stand-alone result -- a new map, also to third order in the coupling constant, but valid exclusively in six dimensions. The map presented here, arrived at by trial and error (starting with an educated guess), is simpler than the one in [9] and highlights a potential ambiguity in the formalism.
\vskip 0.3cm
\ndt Supersymmetric gauge theories, in $D$ dimensions, are characterized by the existence of a Nicolai map $\cT_g$ - of the Yang--Mills fields
\beas
\cT_g \, : \, A_\m^a(x)\ \mapsto\ A_\m^{\prime \, a}(x,g;A) \, , 
\eeas
\vspace*{-0.1cm}
\ndt such that
\begin{itemize}
\item 
The Yang--Mills action without gauge-fixing terms is mapped to the abelian action
\bea
\label{one}
S_0[A']\ =S_g[A]\, ,
\eea
where $S_g[A] = \frac14\int \mathrm{d}x  \, F_{\m\n}^a F_{\m\n}^a$ is the Yang--Mills action 
with gauge coupling $g$ and $F_{\m\n}^a \equiv \pa_\m A_\n^a - \pa_\n A_\m^a + gf^{abc} A_\m^b A_\n^c$
is the field strength.
\item On the gauge surface\footnote{The gauge surface restriction will prove unnecessary for this particular map.} $G^a[A] \equiv \pa^\mu A_\mu^a =0$, the Jacobi determinant of $\cT_g$ is equal to the product of  the MSS and FP determinants, order by order in perturbation theory.
\bea
\label{two}
\det \left( \frac{\d A_\m^{\prime \, a}(x,g;A) }{\d A_\n^b(y)} \right)\ =\ \D_{MSS} [A]\ \D_{FP}[A]\, .
\eea
\item The gauge fixing function
\bea
\label{three}
{\mbox {$G^a[A]$ is a fixed point of $\cT_g$.}}
\eea
\end{itemize}
\ndt A new expression for $A_\mu^{\prime \, a}(x)$ up to order $g^3$ is presented in this paper and shown to satisfy all three requirements above {\it {only}} in $D=6$. 
\vskip 0.3cm
\ndt We work in Euclidean space using the Landau gauge
\be
G^a[A_\mu] = \pa^\mu A_\mu^a\, .
\ee
The results presented below may be adapted to other gauges (the light-cone gauge being of particular interest given potential links to~\cite{SA1}). The free scalar propagator is ($\Box \equiv \pa^\mu \pa_\mu$)
\be
C(x) = \int \frac{d^D k}{(2\pi)^D} \frac{e^{ikx}}{k^2} \quad \Ra \quad
- \Box C(x) = \d(x)\, .
\ee 
The free fermion propagator is (spinor indices suppressed)
\be
\c^\mu\pa_\mu S_0(x)   = \d (x)   \quad \Ra \quad S_0 (x) = - \c^\mu \pa_\mu C(x)\, ,
\ee
$S_0(x-y) = - S_0(y-x)$. In a gauge-field dependent background
\be
\c^\mu (D_\mu S)^{ab}(x) \equiv \c^\mu \Big[ \d^{ac}\pa_\mu - gf^{acd} A_\mu^d(x) \Big] S^{cb}(x)=\d^{ab} \d(x)\, .
\ee 

\section{Result} 
The new result in this paper is the following explicit expression for $\cT_g$ to $\cO(g^3)$.
\bea
\label{boson}
({\cT_g A)_\mu^a}(x)&=&{A_\mu^a}(x)+g\,f^{abc}\,\int\,dy\;\partial_\lambda\,C(x-y)\,A_\mu^b(y)\,A_\lambda^c(y) \nn\\ \nn\\
&&+\,\frac{3}{2}\,g^2\,f^{abc}\,f^{bde}\,\int\,dy\,dz\;\partial_\rho\,C(x-y)\,A_\sigma^c(y)\,\partial_{[\rho}\,C(y-z)\,A_\mu^d(z)\,A_{\sigma]}^e(z) \nn \\[3.5 mm]
&&+\,\frac{3}{2}\,g^3\,f^{abc}\,f^{bde}\,f^{dmn}\,\int\,dy\,dz\,dw\,\partial_\rho\,C(x-y)\,A_\lambda^c(y)\,{\biggl \{} \nn\\
&&\hspace{2.3cm}+\,\partial_\lambda\,C(y-z)\,A_\sigma^e(z)\,\partial_{[\mu}\,C(z-w)\,A_\rho^m(w)\,A_{\sigma]}^n(w) \nn \\
&&\hspace{2.3cm}+\,\partial_\mu\,C(y-z)\,A_\sigma^e(z)\,\partial_{[\sigma}\,C(z-w)\,A_\rho^m(w)\,A_{\lambda]}^n(w) \nn \\
&&\hspace{2.3cm}+\,\partial_\rho\,C(y-z)\,A_\sigma^e(z)\,\partial_{[\sigma}\,C(z-w)\,A_\lambda^m(w)\,A_{\mu]}^n(w) \nn {\biggr \}} \nn\\[3.5 mm]
&&-\,g^3\,f^{abc}\,f^{bde}\,f^{dmn}\,\int\,dy\,dz\,dw\,\partial_\rho\,C(x-y)\,A_\lambda^c(y)\,{\biggl \{} \nn \\
&&\hspace{2.3cm}+\,\partial_\sigma\,C(y-z)\,A_\sigma^e(z)\,\partial_{[\mu}\,C(z-w)\,A_\lambda^m(w)\,A_{\rho]}^n(w) \nn \\
&&\hspace{2.3cm}+\,\partial_\sigma\,C(y-z)\,A_\rho^e(z)\,\partial_{[\sigma}\,C(z-w)\,A_\lambda^m(w)\,A_{\mu]}^n(w) \nn \\
&&\hspace{2.3cm}+\,\partial_\sigma\,C(y-z)\,A_\mu^e(z)\,\partial_{[\sigma}\,C(z-w)\,A_\rho^m(w)\,A_{\lambda]}^n(w) \nn \\ 
\label{bosong^3}
&&\hspace{2.3cm}+\,\partial_\sigma\,C(y-z)\,A_\lambda^e(z)\,\partial_{[\mu}\,C(z-w)\,A_\rho^m(w)\,A_{\sigma]}^n(w)  {\biggl \}} \, ,
\eea

\ndt where $[\mu \nu \rho]=\frac{1}{6}[\mu\nu\rho-\mu\rho\nu+\nu\rho\mu-\nu\mu\rho+\rho\mu\nu-\rho\nu\mu]$. 
\vskip 0.3cm
\ndt It is important to note that this result differs from the one in~\cite{Nic4}. All terms above have the base structure $\displaystyle \partial CA\;\partial CA\;\partial CAA$ at $\mathcal{O}(g^3)$, while the result in~\cite{Nic4} also includes the structures $\displaystyle \partial C\;\partial CAA\;\partial CAA$, $\displaystyle A\;CA\;\partial CAA$ and $\partial C \,\partial (A C)\, A \,\partial C A A$.
\vskip 0.1cm
\ndt Further, terms that overlap with those in~\cite{Nic4}, appear here with different coefficients. As a consequence, the expression above is {\it {not a subset}} of the result in~\cite{Nic4}. 
\vskip 0.1cm
\ndt Finally, while the result in~\cite{Nic4} was valid in all the critical dimensions, we will see that the result in (\ref {boson}) constitutes a map {\bf {only in six dimensions}}.
\vskip 0.3cm
 
\section{Checks of the Result}
\ndt In this section, we prove that expression in (\ref {boson}) satisfies all three requirements, (\ref {one}), (\ref {two}) and (\ref {three}), necessary for it to be a map. The calculations up to $\cO(g^2)$ are identical to those in~\cite{Nic3, Nic4}, so the focus here will be on $\mathcal{O}(g^3)$.

\vskip 0.3cm

\subsection{Gauge condition}

\ndt We begin with the third requirement, listed in (\ref {three}). We need to show that $\pa_\m A_\m^{\prime \, a}(x)  = \pa_\m A_\m^a(x) + \cO(g^4)$. 
\vskip 0.3cm
\ndt We apply $\pa_\m$ to the terms of order $g^3$ in \eqref{boson}. This gives us a symmetric $\pa_\m\,\pa_\rho$ at the beginning of the expression so we eliminate all terms that are anti-symmetric under the exchange $\mu\leftrightarrow\rho$ and find
\bea
\pa_\m A_\m^{\prime \, a}(x) \big\vert_{\cO(g^3)}\!\!\!&=&\!\!\!\frac{3}{2}\,g^3\,f^{abc}\,f^{bde}\,f^{dmn}\,\int dy\,dz\,dw\,\partial_\mu\partial_\rho\,C(x-y)\,A_\lambda^c(y)\,{\biggl \{} \nn\\
&&\hspace{2.3cm}+\,\partial_\mu\,C(y-z)\,A_\sigma^e(z)\,\partial_{[\sigma}\,C(z-w)\,A_\rho^m(w)\,A_{\lambda]}^n(w) \nn \\
&&\hspace{2.3cm}+\,\partial_\rho\,C(y-z)\,A_\sigma^e(z)\,\partial_{[\sigma}\,C(z-w)\,A_\lambda^m(w)\,A_{\mu]}^n(w) {\biggl \}}\nn \\ 
&&-\,g^3\,f^{abc}\,f^{bde}\,f^{dmn}\,\int dy\,dz\,dw\,\partial_\mu\partial_\rho\,C(x-y)\,A_\lambda^c(y)\,{\biggl \{}\nn \\ [3 mm]
&&\hspace{2.3cm}+\,\partial_\sigma\,C(y-z)\,A_\rho^e(z)\,\partial_{[\sigma}\,C(z-w)\,A_\lambda^m(w)\,A_{\mu]}^n(w) \nn \\
&&\hspace{2.3cm}+\,\partial_\sigma\,C(y-z)\,A_\mu^e(z)\,\partial_{[\sigma}\,C(z-w)\,A_\rho^m(w)\,A_{\lambda]}^n(w) {\biggl \}} \, . 
\eea
The first two terms cancel each other under the interchange of $\mu$ and $\rho$. Similarly, the other two terms also cancel out confirming that
\bea
\pa_\m A_\m^{\prime \, a}(x)  = \pa_\m A_\m^a(x) + \cO(g^4) \, .
\eea

\vskip 0.3cm

\subsection{Free Action}
\ndt We now move to the first requirement in (\ref {one}) which states that the transformed gauge field must satisfy
\begin{align}
\frac{1}{2} \int dx\,A_\m^{\prime \, a}(x) \left(- \Box \, \d_{\m\n} + \pa_\m \pa_\n
\right) A_\n^{\prime \, a}(x) \,=\,  \frac{1}{4} \int  dx\,F_{\m \n}^a(x) F_{\m \n}^a(x)  
\,+ \, \cO(g^4) \, . 
\label{eq:FreeAction}
\end{align}
\ndt Because of the invariance of the gauge function, we ignore the second term on the l.h.s. and the corresponding term on the r.h.s. of this equation~\cite{Nic3}. At third  order, \eqref{eq:FreeAction} has two contributions 
\begin{align}
0 &\stackrel{!}{=}  \int dx \bigg( A_\m^{\prime \, a}(x) \big\vert_{\cO(g^3)} \, \Box \,  A_\m^{\prime \, a}(x) \big\vert_{\cO(g^0)} +  A_\m^{\prime \, a}(x) \big\vert_{\cO(g^2)}  \, \Box  \, A_\m^{\prime \, a}(x) \big\vert_{\cO(g^1)}\bigg)  \, .
\end{align}
This expression reads
\begin{align}
\begin{aligned}
&\int dx\, \bigg( A_\m^{\prime \, a}(x) \big\vert_{\cO(g^3)} \, \Box \,  A_\m^{\prime \, a}(x) \big\vert_{\cO(g^0)} +  A_\m^{\prime \, a}(x) \big\vert_{\cO(g^2)}  \, \Box  \, A_\m^{\prime \, a}(x) \big\vert_{\cO(g^1)} \bigg) \nn\\ \nn\\
&\q= \frac{3}{2}\,g^3\,f^{abc}\,f^{bde}\,f^{dmn}\,\int\,dx\,dy\,dz\,dw\,\partial_\rho\,C(x-y)\,A_\lambda^c(y)\,{\biggl \{} \nn\\
&\hspace{2.3cm}+\,\partial_\lambda\,C(y-z)\,A_\sigma^e(z)\,\partial_{[\mu}\,C(z-w)\,A_\rho^m(w)\,A_{\sigma]}^n(w)  \Box \, A_\m^a(x)\nn \\
&\hspace{2.3cm}+\,\partial_\mu\,C(y-z)\,A_\sigma^e(z)\,\partial_{[\sigma}\,C(z-w)\,A_\rho^m(w)\,A_{\lambda]}^n(w)  \Box \, A_\m^a(x)\nn \\
&\hspace{2.3cm}+\,\partial_\rho\,C(y-z)\,A_\sigma^e(z)\,\partial_{[\sigma}\,C(z-w)\,A_\lambda^m(w)\,A_{\mu]}^n(w) \Box \, A_\m^a(x)  {\biggl \}}\nn \\ 
&\qq -\,g^3\,f^{abc}\,f^{bde}\,f^{dmn}\,\int\,dx\,dy\,dz\,dw\,\partial_\rho\,C(x-y)\,A_\lambda^c(y)\,{\biggl \{}\nn \\
&\hspace{2.3cm}+\,\partial_\sigma\,C(y-z)\,A_\sigma^e(z)\,\partial_{[\mu}\,C(z-w)\,A_\lambda^m(w)\,A_{\rho]}^n(w)  \Box \, A_\m^a(x)\nn \\
&\hspace{2.3cm}+\,\partial_\sigma\,C(y-z)\,A_\rho^e(z)\,\partial_{[\sigma}\,C(z-w)\,A_\lambda^m(w)\,A_{\mu]}^n(w) \Box \, A_\m^a(x) \nn \\
&\hspace{2.3cm}+\,\partial_\sigma\,C(y-z)\,A_\mu^e(z)\,\partial_{[\sigma}\,C(z-w)\,A_\rho^m(w)\,A_{\lambda]}^n(w) \Box \, A_\m^a(x)\nn\\
&\hspace{2.3cm}+\,\partial_\sigma\,C(y-z)\,A_\lambda^e(z)\,\partial_{[\mu}\,C(z-w)\,A_\rho^m(w)\,A_{\sigma]}^n(w)  \Box \, A_\m^a(x) {\biggl \}} \nn\\
&\qq +\,\frac{3}{2}\,g^3\,f^{abc}\,f^{bde}\,\int\,dx\,dy\,dz\,dw\;\partial_\rho\,C(x-y)\,A_\lambda^c(y)\,\,\partial_{[\rho}\,C(y-z)\,A_\mu^d(z)\,A_{\lambda]}^e(z) \nn \\
&\hspace{3cm}  \times \, \Box \, \left( f^{a m n}  \pa_\s C(x-w) A_\m^{m}(w) A_\s^{n}(w) \right) \, .
\end{aligned}
\end{align}
We simplify the r.h.s. to obtain
\begin{align}
\begin{aligned}
&\q= \frac{3}{2}\,g^3\,f^{abc}\,f^{bde}\,f^{dmn}\,\int\,dx\,dz\,dw\,\partial_\rho\,A_\m^a(x)\,A_\lambda^c(x)\,{\biggl \{} \nn\\
&\hspace{2.3cm}+\,\partial_\lambda\,C(x-z)\,A_\sigma^e(z)\,\partial_{[\mu}\,C(z-w)\,A_\rho^m(w)\,A_{\sigma]}^n(w) \nn \\
&\hspace{2.3cm}+\,\partial_\mu\,C(x-z)\,A_\sigma^e(z)\,\partial_{[\sigma}\,C(z-w)\,A_\rho^m(w)\,A_{\lambda]}^n(w)\nn \\
&\hspace{2.3cm}+\,\partial_\rho\,C(x-z)\,A_\sigma^e(z)\,\partial_{[\sigma}\,C(z-w)\,A_\lambda^m(w)\,A_{\mu]}^n(w) {\biggl \}}\nn \\ 
&\qq-\,g^3\,f^{abc}\,f^{bde}\,f^{dmn}\,\int\,dx\,dz\,dw\,\partial_\rho A_\m^a(x)\,A_\lambda^c(x)\,{\biggl \{}\nn \\
&\hspace{2.3cm}+\,\partial_\sigma\,C(x-z)\,A_\sigma^e(z)\,\partial_{[\mu}\,C(z-w)\,A_\lambda^m(w)\,A_{\rho]}^n(w) \nn \\
&\hspace{2.3cm}+\,\partial_\sigma\,C(x-z)\,A_\rho^e(z)\,\partial_{[\sigma}\,C(z-w)\,A_\lambda^m(w)\,A_{\mu]}^n(w)  \nn \\
&\hspace{2.3cm}+\,\partial_\sigma\,C(x-z)\,A_\mu^e(z)\,\partial_{[\sigma}\,C(z-w)\,A_\rho^m(w)\,A_{\lambda]}^n(w) \nn\\
&\hspace{2.3cm}+\,\partial_\sigma\,C(x-z)\,A_\lambda^e(z)\,\partial_{[\mu}\,C(z-w)\,A_\rho^m(w)\,A_{\sigma]}^n(w) {\biggl \}} \nn\\
&\qq +\,\frac{3}{2}\,g^3\,f^{abc}\,f^{bde}f^{a m n}\,\int\,dx\,dz\,dw \nn\\
&\hspace{2.6cm}A_\lambda^c(x)\,\,\partial_{[\rho}\,C(x-z)\,A_\mu^d(z)\,A_{\lambda]}^e(z) \partial_\rho\pa_\s C(x-w) A_\m^{m}(w) A_\s^{n}(w) \, .
\end{aligned}
\end{align}
\ndt This is further simplified with some re-writing [for example, $\partial_\rho A_\mu^a(x)  A_\lambda^c(x)$ $\rightarrow$ $\frac{1}{2} \pa_\rho\left(  A_\mu^a(x)  A_\lambda^c(x) \right)$ based on the symmetries $a \leftrightarrow c$ and $\m \leftrightarrow \lambda$]. The r.h.s. simplifes to 
\bea
=\frac{3}{4}\,g^3 f^{abc}f^{bde}f^{dmn}A_{\mu}^a(x)A_{\lambda}^{c}(x)A_{\sigma}^{e}(x) \partial_{\,[\,\sigma} C(x-w) A_{\lambda}^{m}(w) A_{\mu\,]\,}^{n}(w)\, .
\eea
There is a symmetry to these terms: the $\partial CAA$ blocks are invariant under a cyclic permutation of the Lorentz indices. This motivates re-writing the term as
\bea
\label{jacob}
&\frac{1}{4}\,g^3 f^{abc}f^{bde}f^{dmn}\left[A_{\mu}^a(x)A_{\lambda}^{c}(x)A_{\sigma}^{e}(x)+A_{\sigma}^a(x)A_{\mu}^{c}(x)A_{\lambda}^{e}(x)+A_{\lambda}^a(x)A_{\sigma}^{c}(x)A_{\mu}^{e}(x) \right]\nn \\
&\times\; \partial_{\,[\,\sigma} C(x-w) A_{\lambda}^{m}(w) A_{\mu\,]\,}^{n}(w)\\
&=\frac{1}{4}\,g^3 \left[f^{abc}f^{bde}+f^{eba}f^{bdc}+f^{cbe}f^{bda} \right]f^{dmn}A_{\mu}^a(x)A_{\lambda}^{c}(x)A_{\sigma}^{e}(x)\nn \\
&\times\; \partial_{[\sigma} C(x-w) A_{\lambda}^{m}(w) A_{\mu]}^{n}(w)\, .\nn
\eea
We now find, for the first time in this computation, that for (\ref {jacob}) to vanish we need to invoke the Jacobi identity
\bea
f^{abc}f^{bde}+f^{eba}f^{bdc}+f^{cbe}f^{bda}=0\, .
\eea
Thus \eqref{eq:FreeAction} holds up to $\cO(g^3)$.
\vskip 0.5cm
\subsection{Jacobians, fermion and ghost determinants }
\vskip 0.3cm
\ndt Finally, we turn to (\ref {two}), the second requirement. This is, in some sense, the most constraining of the three requirements, demanding that the bosonic Jacobian determinant equal the product of the MSS and FP determinants. Again, this check up to $\mathcal{O}(g^2)$ was performed in \cite{Nic1,Nic3} allowing us to concentrate here on $\mathcal{O}(g^3)$.
\begin{align}
\log \det \left( \frac{\d A_\m^{\prime \, a}(x)}{\d A_\n^b(y)} \right) \bigg\vert_{\cO(g^3)} \stackrel{!}{=} \log \left(\D_{MSS}[A] \ \D_{FP}[A] \right) \bigg\vert_{\cO(g^3)} \, .
\label{Jacob}
\end{align}
It is this non-trivial requirement which results in a dimensional dependence. We prove that the map in (\ref {boson}) satisfies (\ref {Jacob}) only for $D=6$.
\vskip 0.5cm

\subsubsection*{Fermion determinant}

\ndt To compute the fermion determinant, we need to evaluate the following quantity

\bea
\det \left[\gamma^{\mu} \left(\delta^{ab}\partial_{\mu}-gf^{abm}A_{\mu}^m \right) \right]=\det \slashed{\partial}\cdot \det (1-Y)\, ,
\eea
where the relevant functional matrix reads
\bea
Y^{ab}(x,y;A)\,=\,g\,f^{abm} \gamma^\mu \gamma^\nu \partial_\mu C(x-y)\,A_\nu^m(y)\, .
\eea
We use 
\bea
\log \det \big(1-Y\big) = {\rm Tr}\, \log \big(1-Y \big) = -\sum_{n=1}^\infty \frac1{n} {\rm Tr}\, Y^n\, ,
\eea
to arrive at the following five independent terms at order $g^3$
\bea
&&g^3\,f^{abm}\,f^{bcn}\,f^{cap}\,\int\,dx\,dy\,dz\ {\biggl \{} \nn\\[2 mm]
&&\hspace{2.3cm}-\,r\; \partial_\rho\,C(x-y)\,A_\rho^m(y)\,\partial_\lambda\,C(y-z)\,A_\sigma^n(z)\,\partial_\lambda\,C(z-x)\,A_\sigma^p(x) \nn \\
&&\hspace{2.3cm}+\, \displaystyle \frac{r}{3}\;\partial_\rho\,C(x-y)\,A_\lambda^m(y)\,\partial_\lambda\,C(y-z)\,A_\sigma^n(z)\,\partial_\sigma\,C(z-x)\,A_\rho^p(x) \nn \\ 
&&\hspace{2.3cm}+\,\displaystyle \frac{r}{2}\;\partial_\rho\,C(x-y)\,A_\lambda^m(y)\,\partial_\lambda\,C(y-z)\,A_\rho^n(z)\,\partial_\sigma\,C(z-x)\,A_\sigma^p(x) \nn \\
&&\hspace{2.3cm}-\,\displaystyle \frac{r}{6}\;\partial_\rho\,C(x-y)\,A_\lambda^m(y)\,\partial_\sigma\,C(y-z)\,A_\rho^n(z)\,\partial_\lambda\,C(z-x)\,A_\sigma^p(x) \nn \\
&&\hspace{2.3cm}+\,\displaystyle \frac{r}{2}\; \partial_\rho\,C(x-y)\,A_\lambda^m(y)\,\partial_\sigma\,C(y-z)\,A_\rho^n(z)\,\partial_\sigma\,C(z-x)\,A_\lambda^p(x){\biggr \}} \, ,
\eea
where $r$ represents the number of spinor components.

\vskip 0.3cm

\subsubsection*{Ghost determinant}
\ndt For the ghost determinant, we compute
\bea
\det (D_{\mu}\partial^{\mu})= \det\left(\left[\delta^{ab}\partial_{\mu}-gf^{abm}A_{\mu}^m \right]\partial^{\mu} \right)=\det(\Box) \cdot \det(1-X)\, ,
\eea
\ndt where
\bea
X^{ab}(x,y;A) \,=\, g f^{abm}\partial_\mu  C(x-y) A_\mu^m(y)\, .
\eea
Up to $\cO(g^3)$ this yields
\bea
\label{ghost}
&&+\frac{1}{3}\,g^3\,f^{abm}\,f^{bcn}\,f^{cap}\,\int\,dx\,dy\,dz\ \nn \\
&&\hspace{2cm}\partial_\rho\,C(x-y)\,A_\rho^m(y)\,\partial_\lambda\,C(y-z)\,A_\lambda^n(z)\,\partial_\sigma\,C(z-x)\,A_\sigma^p(x)\, .
\eea
\vskip 0.3cm

\subsubsection*{Bosonic Jacobian}
\label{label}

\ndt At $\cO(g^3)$ the logarithm of the Jacobian determinant schematically consists of three terms  
\begin{align}
\begin{aligned}
\label{jacobian}
\log \det \left( \frac{\d A^{\prime \, a}_\m(x)}{\d A_\n^b(y)} \right) \bigg\vert_{\cO(g^3)} 
&=\;  {\rm Tr} \left[ \frac{\d A^\prime}{\d A} \bigg\vert_{\cO(g^3)} \right]  \,- \,
\left( 2 \cdot \frac{1}{2} \right)  {\rm Tr} \left[ \frac{\d A^\prime}{\d A}  \bigg\vert_{\cO(g^2)} 
\frac{\d A^\prime}{\d A} \bigg\vert_{\cO(g^1)} \right] \\[1mm]
&\hspace{2cm}  +  \, \frac{1}{3} \,  {\rm Tr} \left[ \frac{\d A^{\prime}}{\d A} \bigg\vert_{\cO(g^1)} 
\frac{\d A^{\prime}}{\d A} \bigg\vert_{\cO(g^1)} 
\frac{\d A^{\prime}}{\d A} \bigg\vert_{\cO(g^1)} \right]\, ,
\end{aligned}
\end{align}
and the final trace involves setting $\mu=\nu, a=b, x=y$ and integrating over $x$. 
\vskip 0.1cm
\ndt All terms at $\cO(g^3)$ are of the form $\partial CA\,\partial CA\,\partial CAA$. The functional derivative on the very first field, in this structure, vanishes trivially~\cite{Nic3}. The functional differentiation of the field in the middle block produces the structure $\partial CA\,\partial C\,\partial CAA$ not seen elsewhere. These terms vanish as described in the appendix.  Functional differentiation of either field from the last block produces terms with the same structure as those from the fermion and ghost contributions. The table below offers a summary of the various contributions to the Jacobian from~(\ref{jacobian}).
\subsubsection*{Jacobian table}
In the table, colums $2-5$ capture bosonic contributions, summed up in column $6$. Column $7$ contains the sums of the fermion and ghost contributions. The detailed breakdown for the bosonic contributions is as follows: Column 2 contains the contributions from $\cO(g)$ terms when ``cubed". Column 3 lists contributions from $\cO(g)\times \cO(g^2)$. Column 4 has contributions from the $9$ terms in the bosonic result (first three lines of $\cO(g^3)$ from (\ref {boson})). In column 5, we present contributions from the next four lines of (\ref {boson}) ($12$ terms).

\begin{table}[H]
 \centering
\begin{tabular}{|c|c|c|c|c|c|c|}
\hline
&&&&&& \\
Group & $ {(g)}^3$ & $ (g)\times (g^2) $ & 9 Terms & 12 Terms & Boson & MSS+FP \\
&&&&&& \\
\hline
\hline
&&&&&&\\
1 & 0& $\frac{1-D}{2}$ & $\frac{5-2D}{2}$ & $\frac{2}{3}\left(3-D \right)$& $\frac{30-13D}{6}$& $-r$  \\
&&&&&& \\
\hline
&&&&&&\\
2 & $\frac{D-3}{3}$ & $\fr{2}$ &$\frac{D-3}{2}$ & 0& $\frac{5D-12}{6}$ & $\frac{r+1}{3}$  \\
&&&&&& \\
\hline
&&&&&&\\
3 & 1 & $\frac{D-3}{2}$ & $\fr{2}$&$\frac{D-3}{3}$ & $\frac{5D-6}{6}$ & $\frac{r}{2}$  \\
&&&&&& \\
\hline
&&&&&&\\
4 & $-\fr{3}$ & 0 & 0 & $\frac{3-D}{3}$& $\frac{2-D}{3}$ & $-\frac{r}{6}$ \\
&&&&&& \\
\hline
&&&&&&\\
5& 0 &$\frac{1}{2}$ & $\frac{D-3}{2}$& $\frac{2D-6}{3}$ & $\frac{7D-18}{6}$ & $\frac{r}{2}$  \\
&&&&&& \\
\hline
\end{tabular}
\end{table}
\ndt In column $7$, we now set~\cite{Nic3} 
\bea
\label{rD}
r=2(D-2)\, .
\eea 
\vskip 0.3cm
\ndt The {\it {main result}} is that Columns 6 and 7 are equal {\it {only for}} $D=6$.

\vskip 0.3cm

\ndt This completes our proof of (\ref {one}), (\ref {two}) and (\ref {three}). It is curious that we have not had to invoke the gauge condition, which was needed in~\cite{Nic4}, in this proof.

\section{A potential algorithm to generate the map to third order and beyond}

In this section we outline an algorithmic approach to determining the map $T_g$. This involves perturbaively generating higher order expressions in a manner reminiscent of that in~\cite{DL1}. However, the approach presented here comes with the potential advantage of leading to the map directly as opposed to generating the inverse map $T^{-1}_g$.
\vskip 0.3cm
\ndt As mentioned already below equation (\ref {boson}), the structure of the map presented in this paper is simpler than that in~\cite{Nic4}. The entire map in (\ref {boson}), at order $g^3$, involves a single structure. We present below an algorithm that generates exactly this structure suggesting a simple all-order generalization of our results.
\vskip 0.3cm
\ndt We start by noting that the ``base" structure - the order $g$ result - has the form: $\partial C A A$. Our claim is that there exists a realization of the map $T_g$, to all orders, generated entirely by linking a series of $\partial C A$ factors to this base structure.
\vskip 0.3cm
\ndt We illustrate this first at order $g^2$. The map, at this order, would necessarily involve one $\partial C A$ block in addition to the base structure.
\beas
\displaystyle \cT (g^2)\,\sim\,g^2 \;\; \partial CA \;\; \partial CAA\ .
\eeas
We are now guided by the following algorithm. 
\begin{itemize}
\item Sprinkle Lorentz indices on the base  $\partial CAA$ block, such that the indices are all distinct. A set of three terms having the same ``external" structure but with the three indices on the base-block permuted cyclically constitute a ``triplet". 
\item Choose the two Lorentz indices on the first ``block" to be different, for example without loss of generality we can choose them to be $\rho$ and $\lambda$ respectively.
\item Discard all terms with $\mu$ on the $\partial$ of the first block (Note that $A_\mu\,\Box$ acting on such terms, from the left, would trivially vanish).
\end{itemize}
Focussing on order $g^2$
\bea
\displaystyle \cT (g^2)\,\sim\,g^2 \;\; \partial CA \;\; \partial CAA\ ,
\eea
we distribute Lorentz indices on the $\partial CAA$ block. We have three sets of indices at this order: $\rho$, $\lambda$ which are summed over and $\mu$ which is the free index. There is only one ``triplet" possible at this order, with $\mu$, $\rho$ and $\lambda$ all sprinkled on the last block. Hence, at this order, the algorithm generates three terms in the map $T_g$:
\bea
T_g: g^2\; \partial_{\rho}CA_{\lambda}\;\partial_{[\rho}CA_{\lambda}A_{\mu]}\ . 
\eea
Moving to order $g^3$, our procedure asks that we add two $\partial C A$ structures to the base structure. So we have
\bea
\displaystyle O(g^3)=g^3 \;\; \partial CA \;\; \partial CA \;\; \partial CAA\ .
\eea
We again distribute Lorentz indices on the $\partial CAA$ block. At this order, we have four sets of indices to work with: $\rho$, 
$\sigma$, $\lambda$ all summed over and $\mu$ which is free. There are $4$ ways of selecting $3$ different indices (triplets) from the available set. Without loss of generality we choose the Lorentz indices on the first block to be $\rho$ and $\lambda$ respectively. This leaves us with two indices and two slots, which is two arrangements for each triplet, except for one, where we have the same index ($\sigma$ in this convention), and hence only one arrangement. This gives us seven triplets, or $21$ terms at order $g^3$, and the map

\bea
T_g: g^3 &\partial_{\rho}CA_{\lambda}\partial_{\lambda}CA_{\sigma}\partial_{[\mu}CA_{\rho}A_{\sigma]}\nn \\
&\partial_{\rho}CA_{\lambda}\partial_{\sigma}CA_{\lambda}\partial_{[\mu}CA_{\rho}A_{\sigma]}\nn \\
& \partial_{\rho}CA_{\lambda}\partial_{\mu}CA_{\sigma}\partial_{[\sigma}CA_{\rho}A_{\lambda]}\nn \\
& \partial_{\rho}CA_{\lambda}\partial_{\sigma}CA_{\mu}\partial_{[\sigma}CA_{\rho}A_{\lambda]}\nn \\
& \partial_{\rho}CA_{\lambda}\partial_{\rho}CA_{\sigma}\partial_{[\sigma}CA_{\lambda}A_{\mu]}\nn \\
& \partial_{\rho}CA_{\lambda}\partial_{\sigma}CA_{\rho}\partial_{[\sigma}CA_{\lambda}A_{\mu]}\nn \\
& \partial_{\rho}CA_{\lambda}\partial_{\sigma}CA_{\sigma}\partial_{[\mu}CA_{\lambda}A_{\rho]}\ ,
\eea
exactly matching the structures that appear in (\ref {boson}). We note that while this algorithm does not determine the overall constants, it does generate the terms in sets that conveniently satisfy the gauge constraint. It is fairly straightforward to write down the structures expected at order $g^4$ although performing the relevant checks (particularly of the determinants) is technically more involved.

\begin{center}
* ~ * ~ *
\end{center}

\ndt We conclude that (\ref {boson}) represents an alternate Nicolai map~\cite{OL} in six dimensions, up to $\cO(g^3)$, distinct from the map in~\cite{Nic4}. This raises the possibility that there exists a dimension-dependent map that differs for each critical dimension. However, we note that the checks to this order for this particular map do not guarantee that this map will work at next/higher order\footnote{If this map survives to higher orders, the gauge condition may become necessary, in keeping with~\cite{Nic4}.}. The result in~\cite{Nic4} is different because it is derived from the R-prescription and is limited to $\cO(g^3)$ only because the procedure becomes technically involved at higher orders. We note here that ambiguities in constructing Nicolai maps have been previously flagged and discussed in~\cite{NotU}. In this context, the role of the light-cone gauge is an interesting issue we hope to revisit.
\vskip 0.1cm
\ndt There is a third and rather unlikely outcome: that six dimensions is special for yet unknown reasons. For another curious result within this formalism that singles out six dimensions, see equation (3.10) in~\cite {Nic5}. $D=6$ is also home to the mysterious $\mathcal N=(2,0)$ theory~\cite{N=2} which still lacks a complete Lagrangian description~\cite{SA}. 

\vskip 0.5cm
\noindent {\bf Acknowledgments:} We are grateful to Hermann Nicolai for detailed discussions. We thank Olaf Lechtenfeld for comments and correspondence.
\vskip 0.5cm

\appendix
\section{Jacobian Calculation}
\vskip 0.3cm

We present below a part of the calculation referred to below equation (\ref {jacobian}). 

\subsection*{First set of terms at $\cO(g^3)$}
These are the details for the first nine terms in (\ref {boson}).

\subsubsection*{Line $3$ in (\ref {boson})} 
Functional differentiation of the middle block field in each of the first three lines yields
\bea
&&\frac{\delta {A_\mu^a}'(x)}{\delta A_\nu^p(v)}=\frac{1}{2}\,g^3\,f^{abc}\,f^{bde}\,f^{dmn}\,\int\,dy\,dz\,dw\,\partial_\rho\,C(x-y)\,A_\lambda^c(y)\,\partial_\lambda\,C(y-z) \delta_\sigma^\nu\,\delta^{ep}\,\delta(z-v)  \nn \\
&&{\biggl \{}\partial_\mu\,C(z-w)\,A_\rho^m(w)\,A_\sigma^n(w)+\partial_\rho\,C(z-w)\,A_\sigma^m(w)\,A_\mu^n(w)+\partial_\sigma\,C(z-w)\,A_\mu^m(w)\,A_\rho^n(w) {\biggl \}}\, ,\nn
\eea
tracing over here involves setting $\mu=\nu\ $, $a=p$, $x=v$ and integrating over $x$. This is then
\bea
&&\frac{1}{2}\,g^3\,f^{abc}\,f^{bda}\,f^{dmn}\,\int\,dx\,dy\,dw\,\partial_\rho\,C(x-y)\,A_\lambda^c(y)\,\partial_\lambda\,C(y-x)\nn \\ 
&&{\biggl \{}\partial_\mu\,C(x-w)\,A_\rho^m(w)\,A_\mu^n(w)+\partial_\rho\,C(x-w)\,A_\mu^m(w)\,A_\mu^n(w)+\partial_\mu\,C(x-w)\,A_\mu^m(w)\,A_\rho^n(w){\biggl \}}\, .\nn
\eea
The first and third terms above cancel against each other while the middle terms vanishes (symmetry argument) so these three lines do not contribute to the Jacobian trace.
\vskip 0.3cm
\subsubsection*{Line $4$ in (\ref {boson})} 
After Functional differentiation and tracing over we have
\bea
&&\frac{1}{2}\,g^3\,f^{abc}\,f^{bda}\,f^{dmn}\,\int\,dx\,dy\,dw\,\partial_\rho\,C(x-y)\,A_\lambda^c(y)\,\partial_\sigma\,C(y-x)\nn \\
&&{\biggl \{}\partial_\sigma\,C(x-w)\,A_\rho^m(w)\,A_\lambda^n(w)+\partial_\rho\,C(x-w)\,A_\lambda^m(w)\,A_\sigma^n(w)+\partial_\lambda\,C(x-w)\,A_\sigma^m(w)\,A_\rho^n(w) {\biggl \}}\, . \nn 
\eea
Note that $\partial_\sigma^y\,C(y-x)\,=\,-\,\partial_\sigma^x\,C(x-y)$ meaning that the first line above is symmetric in $\rho, \sigma$ while the bracket is anti-symmetric in the same two indices. Hence this contribution vanishes.
\vskip 0.3cm
\subsubsection*{Line $5$ in (\ref {boson})} 
After differentiating and tracing this reads
\bea
&&\frac{1}{2}\,g^3\,f^{abc}\,f^{bda}\,f^{dmn}\,\int\,dx\,dy\,dw\,\partial_\rho\,C(x-y)\,A_\lambda^c(y)\,\partial_\rho\,C(y-x)\nn\\
&&{\biggl \{}\partial_\mu\,C(x-w)\,A_\lambda^m(w)\,A_\mu^n(w)+\partial_\lambda\,C(x-w)\,A_\mu^m(w)\,A_\mu^n(w)+\partial_\mu\,C(x-w)\,A_\mu^m(w)\,A_\lambda^n(w){\biggl \}}\, .\nn
\eea
These three terms vanish by the same arguments that applied to the terms in line $4$ of (\ref {boson}).

\subsection*{Second set of terms at $\cO(g^3)$}
\ndt We have twelve remaining terms in (\ref {boson}). Functional differentiation and trace in the middle block yields
\bea
&&\frac{3-D}{3}\,g^3\,f^{abc}\,f^{bda}\,f^{dmn}\,\int\,dx\,dy\,dw\,\partial_\rho\,C(x-y)\,A_\lambda^c(y)\,\partial_\sigma\,C(y-x)\nn \\
&&{\biggl \{}\partial_\sigma\,C(x-w)\,A_\rho^m(w)\,A_\lambda^n(w)+\partial_\rho\,C(x-w)\,A_\lambda^m(w)\,A_\sigma^n(w)+\partial_\lambda\,C(x-w)\,A_\sigma^m(w)\,A_\rho^n(w) {\biggl \}}\, . \nn 
\eea
These three term vainsh by using $\partial_\sigma^y\,C(y-x)\,=\,-\,\partial_\sigma^x\,C(x-y)$ as  the first line above is symmetric in $\rho, \sigma$ while the bracket is anti-symmetric in the same two indices. So this contribution vanishes.

\end{document}